\documentclass[reprint,
fleqn, amsmath,amssymb, aps,
]{revtex4-1} \usepackage{color} \usepackage{graphicx}
\usepackage{dcolumn}
\usepackage[mathlines]{lineno}

\begin{document}


\title{Quantum transport by spin-polarized edge states in graphene nanoribbons
in the quantum spin Hall and quantum anomalous Hall regimes }

\author{Nezhat Pournaghavi,$^1$ Cecilia Holmqvist,$^1$ Anna Pertsova,$^2$ and C. M.
Canali$^1$}


 
\affiliation{$^1$ Department of Physics and Electrical Engineering, Linnaeus
University, 392 34 Kalmar, Sweden\\ $^2$ Nordita, Roslagstullsbacken 23, SE-106 91
Stockholm, Sweden}


\begin{abstract} 
\noindent Using the
non-equilibrium Green\noindent 's function method and the Keldysh formalism, we
study the effects of spin-orbit interactions and time-reversal symmetry
breaking exchange fields on
non-equilibrium quantum transport in graphene armchair nanoribbons. We identify
signatures of the quantum spin Hall (QSH) and the quantum anomalous Hall (QAH)
phases in non-equilibrium edge transport by calculating
the spin-resolved real space charge density and
local currents at the nanoribbon edges.  We find that
the QSH phase, which is realized in a system with intrinsic spin-orbit coupling,
is characterized by chiral counter-propagating local spin currents summing up to
a net charge flow with opposite spin polarization at the edges. In the QAH
phase, emerging in the presence of Rashba spin-orbit coupling and a
ferromagnetic exchange field, two chiral edge channels with opposite spins 
propagate in the same direction at each edge, 
generating an unpolarized charge current and a quantized
Hall conductance $G = 2 e^2/h$. 
Increasing the intrinsic spin-orbit
coupling causes a transition from the QAH to the QSH phase, evinced
by characteristic changes in the non-equilibrium edge transport. 
In contrast, an antiferromagnetic exchange field can coexist with a QSH phase, but
can never induce a QAH phase due to a symmetry that combines time-reversal and sublattice
translational symmetry.


\end{abstract}

\pacs{73.21.HB, 85.75.Hh.} \maketitle


\section{\label{sec:level1}INTRODUCTION }

Two-dimensional (2D) topological insulators (TIs) in the presence of
time-reversal symmetry 
(TRS)
can display the quantum spin Hall (QSH) effect, and
therefore are also known as QSH insulators~\cite{Qi2010}.  In the QSH state the
electronic system is characterized by a non-trivial topological bulk gap, opened
by spin-orbit interaction (SOI), separating the valence and conduction bands.
In a Hall bar geometry, a QSH insulator hosts at each edge a pair of gapless
one-dimensional (1D) channels, with opposite spin and
moving in opposite directions, whose linear dispersions cross inside the bulk gap.
The charge current carried by these edge states
is dissipationless in the absence of magnetic scattering, and the longitudinal
conductance is quantized in units of $e^2/h$.  At the same time a QSH state
displays spin accumulation at the edges, leading to a spin Hall resistance,
which typically is not quantized, since in general spin is not conserved.  

The first theoretical prediction of the QSH effect was made by Kane and Mele in
2005~\citep{Kane2005a, Kane2005b} on the basis of 2D models describing graphene
nanoribbons. In bulk graphene, the conduction and valence bands meet at two
inequivalent Dirac points (DPs). Intrinsic SOI opens a small topological gap at
these DPs, inside which 1D edge states can appear, causing the QSH effect.  Since the
intrinsic SOI in pure graphene is very weak, the resulting gap is too
small for the QSH effect to be observed (the SOI
induced gap is estimated to be a fraction of meV in
graphene~\cite{min2006,boettger2007}). On the other hand, the QSH effect has been
verified experimentally in mercury tellurride (HgTe) quantum wells, another 2D
TI~\cite{konig2007}.  Nevertheless, since graphene is such a remarkable
opto-electronic 2D material, intensive efforts to realize
experimentally the QSH state have continued in recent years,
focusing on the artificial enhancement of the intrinsic SOI, e.g. by embedding
heavy atoms into the graphene lattice~\cite{Klimovskikh2016}. 

From a fundamental point of view, graphene is the ideal model system to
investigate the fate of the QSH effect in 2D TIs when 
TRS is
broken, e.g. by the proximity of a magnetic substrate inducing an exchange
interaction on the graphene electrons.  A theoretical study~\cite{Yang2011}
based on the Kane and Mele graphene model in the presence of Rashba SOI and a
ferromagnetic (FM) exchange field 
shows that the exchange interaction can drive the system from the QSH regime
into a new topological phase known as the quantum anomalous Hall  (QAH) regime.
Like the ordinary quantum Hall effect, the QAH effect refers to the quantization of
the Hall conductance in a 2D electron system where 
TRS is
broken. In contrast to the quantum Hall effect, the QAH effect occurs in the absence
of an external magnetic field.  The possibility of this effect was first
proposed by Haldane based on studies of a tight-binding toy model on the same
honeycomb lattice in the presence of a periodic
magnetic field with zero net flux~\cite{haldane1988}.

In Ref.~\onlinecite{Yang2011} and in other theoretical work based either on
similar lattice models~\citep{Qiao2010} or on \textit{ab initio} studies of a
graphene sheet on magnetic insulator substrates~\citep{Qiao2014}, the emergence
of the QAH effect in graphene is typically established by computing the Berry
curvature and the Chern number for the valence bands of the 2D bulk system. A
non-zero Chern number corresponds to a quantized Hall conductance.

The purpose of the present paper is to investigate the effect of 
TRS-breaking exchange fields on the QSH phase in graphene directly from the
point of view of out-of-equilibrium quantum transport in a nanoribbon.
Specifically, we carry out a computational study of the non-equilibrium spin-polarized charge density
and local current at the edges of an armchair graphene nanoribbon, which
characterize the 1D conducting states responsible for both the QSH and the
QAH effect.
Typically, transport properties are experimentally determined via conductance measurements obtained using four-probe techniques. However, the spatial details of the current can be measured using scanning probe microscopy~\cite{Topinka2003}, optical Kerr rotation microscopy~\cite{Crooker2005}, or
indirectly by nuclear magnetic resonance techniques~\cite{Walz2014}.
A FM exchange field drives the system into a QAH phase when only a Rashba SOI is present.
However, when both Rashba and intrinsic SOI are present,
we show that there is a regime, controlled by their relative strength,
where the QSH phase survives even in the presence of the FM
exchange field. In contrast, 
an antiferromagnetic (AFM) exchange field can never induce 
a transition from a QSH to a QAH phase, 
due to a combination of 
TRS and sublattice translational
symmetry. 

The paper is organized as follows. In Sec.~\ref{sec:level2}, we
introduce the tight-binding model and the formalism for quantum transport in
graphene nanoribbons.  In particular, the calculation of non-equilibrium
spin-polarized density and the local current is discussed. The results of our
calculations are presented in Sec.~\ref{sec:level3}.  In Sec.~\ref{sec:level31},
we describe spin-polarized transport in the QSH regime.  The effect of an
exchange field on transport and the transition to the QAH state are studied in
Sec.~\ref{sec:level32}.

\section{\label{sec:level2}MODEL AND METHOD} In order to calculate transport properties of a graphene nanoribbon, we use the
non-equilibrium Green's function (NEGF) technique in a two-terminal system consisting
of a central channel attached to two semi-infinite leads. 
The Hamiltonian of the central region can be written in the second quantized form
\citep{Konschuh2010}
\begin{eqnarray}\label{eq:1}  H  &=& -t\sum _{<i,j>,\sigma }c_{i\sigma}^{\dag
}c_{j\sigma} +  M\sum _{i,\sigma } \xi _{i} c_{i\sigma}^{\dag }
{\sigma}_{z}^{\sigma \sigma}c_{i\sigma } \nonumber\\ && + i \frac{ t_{\rm SO}}{\sqrt
3} \sum _{\ll i,j\gg ,\sigma \sigma '}\nu _{ij}  c_{i\sigma }^{\dag }
{\sigma}_{z}^{\sigma \sigma '} c_{j\sigma'} \nonumber  \\
&& +it_{\rm R} \sum _{<i,j>,\sigma \sigma '}  c_{i\sigma}^{\dag }
\left(\boldsymbol{\sigma} \times \hat{d}_{ij} \right)_{z}^{\sigma \sigma}
c_{j\sigma '} .  \end{eqnarray}
Here,  $c_{i,\sigma}^\dag(c_{i,\sigma}^{\phantom \dag})$ is the electron
creation (annihilation) operator with spin $\sigma$ at site $i$ on the honeycomb
lattice and $\boldsymbol{\sigma}$ is the vector of Pauli matrices.  The first term in Eq.~\ref{eq:1} is the usual nearest-neighbor (NN) hopping term with $t=
2.7$~eV. The second term represents an exchange field of strength $M$, ferromagnetic when $\xi_i= 1$ and antiferromagnetic when $\xi_i = \pm
1$, with the sign being opposite on the two sublattices.  The third term,
involving the next nearest-neighbor (NNN) sites $i$ and $j$, is the intrinsic
SOI with strength $t_{\rm SO}$.  The constant $v_{ij}=\pm 1$, depending on
whether the NNN hopping is clockwise or counter clockwise with respect to the
positive $z$ axis.  Finally, the fourth and last term is a spin-dependent NN
hopping representing a Rashba SOI, where $\hat d_{ij}$ is the unit vector in
real space connecting site $i$ to site $j$.

Here, we will consider values of the SOI parameters that are large enough to produce significant effects. This model is therefore supposed to represent graphene nanoribbons where the SOI has been artificially enhanced, or other 2D atomic monolayers on a honeycomb lattice such as stanene~\cite{Zhu2015, Xiong2016, Wang2016}, where intrinsic SOI is considerably larger than in graphene and is expected to produce a large-gap 2D QSH effect.
Given the
Hamiltonian of the central channel, 
the spin-dependent retarded ($r$) and advanced ($a$) Green\noindent 's functions are given by $\boldsymbol{{\cal G}}^{r} (E)=\left[E^{+}
\boldsymbol{I}-\boldsymbol{H}-\boldsymbol{\Sigma}_{L}^{r}(E)-\boldsymbol{\Sigma}_{R}^{r}(E)\right]^{-1}=\left[{\boldsymbol{\cal G}}^{a}(E)\right]^{\dag }$ ~\citep{0521599431}, where $E^+\equiv E + i 0^+$ and $\boldsymbol{I}$ is an identity matrix. In this equation, $\boldsymbol{\Sigma} _{L(R)}^{r} (E)=
\boldsymbol{H}_{L(R),C}^{\dag } \boldsymbol{g}_{L(R)}(E)\boldsymbol{H}_{L(R),C}$
is the self-energy due to the connection of the left (right) semi-infinite
electrode to the central channel. $\boldsymbol{g}_{L(R)}$ is the surface Green's
function of the left(right) lead; $\boldsymbol{g}_{R}$ is calculated iteratively by adding one unit cell from the central region at a time using the method of Ref.~\onlinecite{Sancho1984}.
Therefore, the Green's functions of a central region consisting of $M$ unit cells each containing $N$ atoms can be efficiently calculated using $2 N \times 2N$ matrices, where 2 comes from the spin-degree of freedom.
$\boldsymbol{H}_{L(R),C}$ is
the tunneling Hamiltonian between the central region and the left(right) lead.
The Hamiltonian of the leads, as well as the tunneling
Hamiltonian, is identical to that of the central region, namely it is given by Eq.~\ref{eq:1}, except that it does not include the exchange field.
Now that we have the Green's functions, the conductance can be calculated as
$G_{\sigma \sigma '} (E)=\frac{e^{2} }{h}
\mathrm{Tr} \left[ \boldsymbol{\Gamma} ^{L} (E) \boldsymbol{\cal G}^{r} (E) \boldsymbol{\Gamma}^{R} (E) \boldsymbol{\cal G}^{a} (E)\right]_{\sigma \sigma'}=\frac{e^{2} }{h} T_{\sigma \sigma '}(E)$, 
where
$\boldsymbol{\Gamma}^{L,R} = i[ \boldsymbol{\Sigma}^r_{L/R}- (
\boldsymbol{\Sigma}^r_{L/R} )^\dag]$ is the broadening due to the electrode
contacts and the trace is taken over the sites in the $\sigma \sigma'$ sub-matrix.

In order to calculate the steady-state local transport properties in the full
NEGF formalism, we need both the retarded and lesser Green\noindent 's
functions, which contain information about the density of available states and
how electrons occupy these states, respectively. In the phase-coherent regime
where interaction self-energy functionals are zero, by using the Keldysh
formalism we can simply write the lesser Green\noindent 's function in terms of
the retarded Green\noindent 's function and the broadening matrices as
\citep{Cresti2003}
\begin{eqnarray}\label{eq:2} &&{\cal G}_{i \sigma , j \sigma'}^< (E)
=[\boldsymbol{\cal G}^r (E) \boldsymbol{\Sigma}^{<}_L+\boldsymbol{\Sigma}^{<}_R
\boldsymbol{\cal G}^a (E)]_{i \sigma, j \sigma'} \nonumber \\
&&=i[\boldsymbol{\cal G}^r(E)(f_L(E)\boldsymbol{\Gamma}^{L} + f_R (E)
\boldsymbol{\Gamma}^{R}) \boldsymbol{\cal G}^a(E) ]_{i \sigma , j \sigma'} .
\quad \quad \end{eqnarray}
Once the lesser Green's function are known, we calculate the
out-of-equilibrium, spin-resolved, local charge density, $n_i^\sigma$. Instead of the full non-equilibrium local charge density \cite{Mahfouzi2013}, we use the non-equilibrium definition
\citep{Nikolic2006, Zrbo2007, Chen2010}
\begin{equation}\label{eq:3} n_i^\sigma =\frac{e}{4\pi}\int_{E_{\rm
F}-eV/2}^{E_{\rm F}+eV/2} dE \, {\cal G}_{i\sigma,i \sigma}^< (E)
\end{equation}
as this quantity sensitively depends on the contribution from the edge states appearing in the bulk gap and therefore is useful for investigating the QSH and QAH phases in the transport regime.

To address the transport properties of the nanoribbon, it
is appropriate to use the notion of bond currents \cite{todorov2002}.  This
concept arises naturally from the time derivative of the electron number
operator at site $i$, $N_i=\sum_{\sigma}c^\dagger_{i\sigma}c_{i\sigma}$.  The equation of motion
$d N_i/dt=[N_i,H]/i\hbar$ leads to a continuity equation for the charge given
by $e \, d N_i/dt + \sum_j {\cal J}_{ij}=0$
\cite{Nikolic2006}, 
where ${\cal J}_{ij}$ is the bond charge current operator
describing the charge current flowing from site $i$ to the neighboring site $j$
via a bundle of flow lines connecting the two sites \cite{todorov2002}.  The
bond charge current can be seen as the sum over spin-resolved bond charge currents, ${\cal J}_{ij} = \sum_{\sigma \sigma'} {\cal J}_{ij}^{\sigma
\sigma'}$, where the spin-dependent bond charge current operator is
\begin{equation}\label{eq:4} {\cal J}_{i j}^{\sigma \sigma'}
=\frac{e}{i \hbar} \left( c^\dagger_{j \sigma'} t^{\sigma' \sigma}_{j i} c_{i
\sigma}   - H.c. \right) \end{equation} and $t^{\sigma' \sigma}_{j i}$ includes
all the hopping parameters connecting the two different sites as specified by
the Hamiltonian \ref{eq:1}.  I.e., $t^{\sigma' \sigma}_{j i}$ incorporates the
spin-dependent and spin-independent Rashba NN hopping as well as the NNN
hopping. Taking the quantum-statistical average of Eq.~\ref{eq:4},
the spin-resolved charge current, $J_{i j}^{\sigma \sigma'}=\langle {\cal J}_{i
j}^{\sigma \sigma'} \rangle$, can be expressed in terms of non-equilibrium
Keldysh Green's functions \cite{caroli1971}.  
Neglecting the SOI contributions in $t^{\sigma' \sigma}_{j i}$ because $t_{\rm R},
t_{\rm SO}\ll t$, the NN spin-resolved bond charge current is diagonal in spin space
such that $J_{ij}^{\sigma \sigma}=J_{ij}^\sigma$, and can be written as \cite{Nikolic2006}
\begin{equation} \label{eq:5} J_{ij}^\sigma =\frac{e}{h} t \int_{E_{\rm
F}-eV/2}^{E_{\rm F}+eV/2} dE \, [ {\cal G}_{i\sigma,j \sigma}^< (E) 
\\
- {\cal G}_{j\sigma, i\sigma}^< (E)]. \end{equation} However, note that
  $J_{ij}^\sigma$ still depends on the intrinsic and Rashba SOI via the Green's
functions.

\section{\label{sec:level3}RESULTS AND DISCUSSION}The electronic states of graphene
nanoribbons, and therefore their transport properties, strongly depend on their edge structure. There are two different types of graphene nanoribbons, zigzag
graphene nanoribbons (ZGNRs) and armchair graphene nanoribbons (AGNRs). In the absence of SOI, ZGNRs are always metallic; AGNRs, which are the ones considered here, can be either metallic or semiconducting depending on their
width. An AGNR is metallic for widths such that the number of dimers $N'$ along
the transverse direction in a unit cell consisting of $N = 2N'$ atoms satisfies
the relation $N' = 3M-1$, with $M$ being an integer.  (See Fig.~\ref{figuretwo} for an AGNR with $N= 2N' = 196$.)

\subsection{\label{sec:level31} Intrinsic spin-orbit coupling: the QSH effect}

\begin{figure}[b] \centering
\includegraphics[width=8.8cm,height=10cm,keepaspectratio]{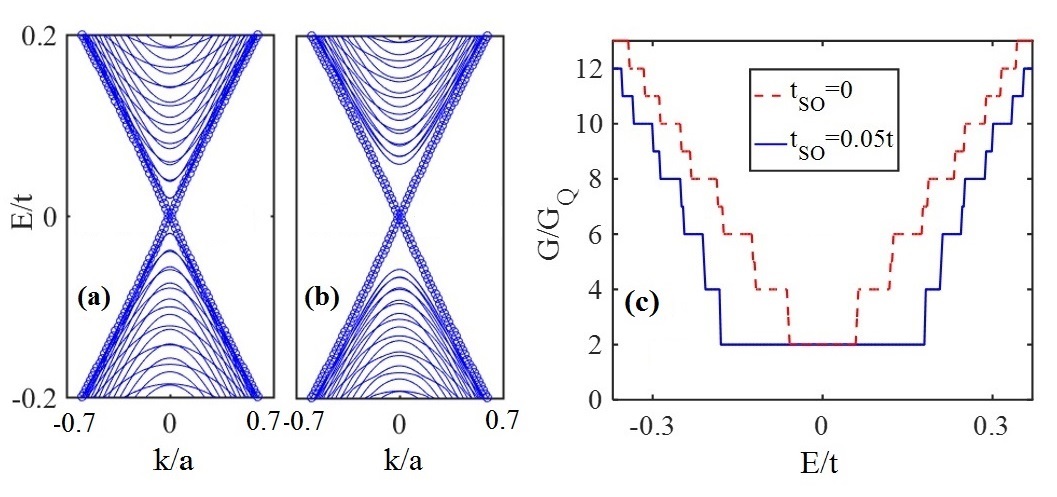}
\caption{\label{figureone} {The band structure of a metallic 196-AGNR (a) without and (b) with  intrinsic spin-orbit coupling. (c) The spin-resolved conductance of the same system.}} \end{figure}

We start  with the case  where only intrinsic SOI is included in the model,
leading to the QSH regime.  Figs.~\ref{figureone}(a) and (b) show the band structure for an $N= 196$ nanoribbon without and with intrinsic SOI, respectively. In both cases two linearly dispersing
bands, crossing at $k=0$ at energy $E= 0$, traverse an energy gap $E_g$ separating all other positive- and negative-energy 1D energy subbands with ordinary quadratic dispersion. When $t_{\rm SO} = 0$ the energy gap $E_g$ shrinks with increasing width, and bulk 1D bands start to pile up toward $E=0$. In contrast, when $t_{\rm SO} \neq 0$ the gap $E_g$
remains finite and converges to the SOI-induced gap of 2D bulk graphene. This feature of the bandstructure has implications for the transport properties:
while in the absence of SOI the energy region where  ${G} = 2 {G}_Q$ ($G_Q= e^2/h$ being the conductance quantum) vanishes for large-width nanoribbons, in the presence of SOI this region remains finite
and is controlled by the SOI strength.  The two branches with linear dispersion are spin-degenerate ``edge states,'' in the sense that they emerge in a graphene system
with edges, such as a nanoribbon. 
However, their nature is quite different, depending on
whether SOI is present or not. In the first case, for a given $k$, the two
states with opposite spin components are entirely space localized at the opposite
edges of the nanoribbon.  Therefore, in the presence of SOI, each edge has a
right-moving ($+k$) state of a given spin and a left-moving ($-k$) state with
opposite spin, both of energy $E = v_{\rm F} k$.  
In contrast, when SOI is
absent, the two states with the same $k$ and opposite spin have the same space wave function, which is distributed throughout 
the nanoribbon. 

These linearly-dispersed edge states completely determine the conductance of
the nanoribbon when the Fermi energy is located within the energy gap $E_g$. As
shown in Fig.~\ref{figureone}(c), the topologically trivial ($t_{\rm SO} = 0$)
and nontrivial ($t_{\rm SO} \neq 0$) state display
the same
transport behavior: the total conductance is quantized, i.e. ${G} = 2 {G}_Q$.
This result is expected, since
in this disorder-free nanoribbon for both cases there are two perfectly conducting
edge states.  

\begin{figure}[t] \centering  
\includegraphics[width=8.5cm,height=10cm,keepaspectratio]{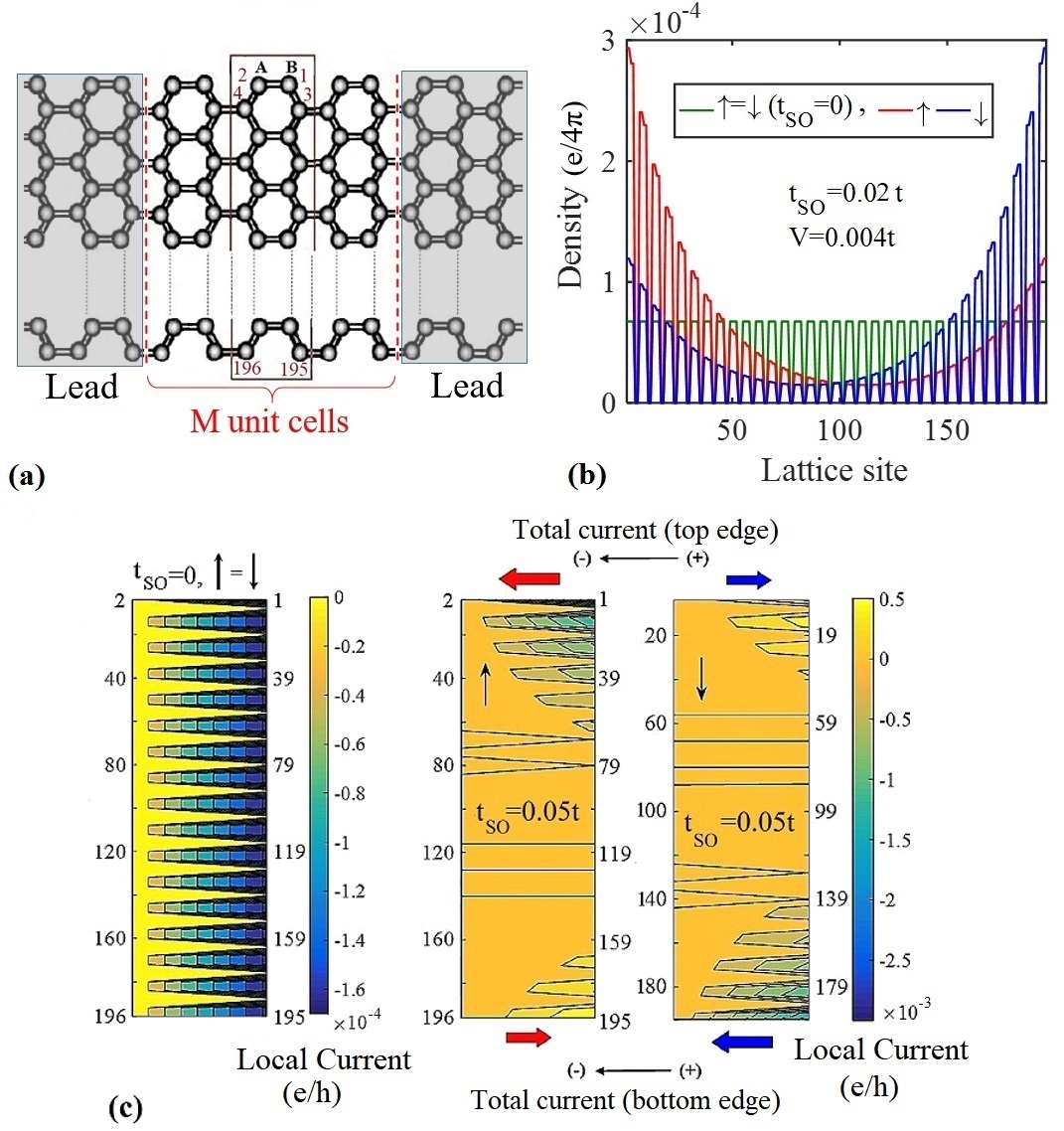}
\caption{\label{figuretwo} (a) Schematic view of the device consisting of a finite central region connected to two semi-infinite leads. The unit cell of an AGNR containing 196 atoms ($\sim 11.9$ nm wide) is shown in the solid box. A and B indicate the sub-lattices and the numbers label the lattice sites within the unit cell. The number of unit cells within the central region is $M=100$ ($\sim 42.4$ nm). (b) Non-equilibrium spin-resolved
charge density of a 196-AGNR in the absence of (green line) and in the presence of
(red and blue lines) of intrinsic SOI as a function of lattice site. (c) 
Local spin-resolved charge current in a unit cell shown in (a) (labels show the atoms' indices). Left panel: SOI is zero. Spin-up and spin-down contributions are identical. Middle and right panels: spin-up and spin-down carrier contributions respectively, when intrinsic SOI is present. The current is now localized at the edges of the nanoribbon. The large red and blue arrows show the direction of the charge current at the edges for spin-up and spin-down electrons, respectively (the length of the arrows is proportional to the current). The black horizontal
arrows show the direction of the net total charge current at the edges.  
} \end{figure}

The onset of the QSH effect in transport becomes evident if we look at the out-of-equilibrium, spin-resolved
charge density in a device made by an AGNR (see Fig.~\ref{figuretwo}(a)),
plotted as a function of lattice sites shown in Fig.~\ref{figuretwo}(b). Here,
the charge density is computed using Eq.~\ref{eq:3} with $E_{\rm F} = 0$ and $eV
= 0.004t$. As can be seen in Fig.~\ref{figuretwo}(b),
in the steady-state regime, states with
energy $|E|  \le eV/2$ contributing to the out-of-equilibrium density are
predominately localized at the edges of the nanoribbons.  Their net occupancy at
a given position is not the same for spin-up and spin-down states, generating a
spin accumulation of opposite sign at the edges of the nanoribbon.  This
non-trivial spin dependence is totally absent when $t_{\rm SO} = 0$ (see the green line in  Fig.~\ref{figuretwo}(b)).

Further evidence of the QSH regime in transport is provided by the local
spin-resolved charge current, shown in Fig.~\ref{figuretwo}(c).  More precisely,
here we plot the bond current (calculated with Eq.~\ref{eq:5}) flowing through
the 
horizontal dimers in the nanoribbon unit cell. 
The color scale is used to indicate both the direction and the intensity of the bond currents through the horizontal dimers of one unit cell (see Fig.~\ref{figuretwo}(a) for the labeling of the dimer atoms). The apparent horizontal variation is related to the way we plot the intensity of this bond current for a given dimer, by inserting the total value of the current at one atom and going smoothly to zero to the other one. The unit cell is then repeated unchanged throughout the central part of the device.
The potential drop of $eV = 0.004t$ induces a net charge current from right to left.
In the absence of SOI (left panel in Fig.~\ref{figuretwo}(c)), the bond current
is the same for spin-up and spin-down electrons and is rather homogeneously
distributed throughout the nanoribbon. In contrast, when SOI  is present, the
bond current is mainly concentrated at the edges, where spin-up and spin-down
electrons move in opposite directions.  For example, at the top edge  spin-up
electrons (middle panel) move from  right to left. At the same edge spin-down
electrons (right panel) move from left to right, but their bond current is
smaller than the bond current of the spin-up states, resulting in a net
spin(-up)-polarized charge current from right to left. Similarly at the bottom
edge we have a net  spin(-down)-polarized charge current also flowing from right
to left. Thus, we have a net bond charge current, localized at the two edges of
the nanoribbon, where it is spin-polarized in opposite directions.
Together with the quantization of the conductance and the spin accumulation at the edges,
these properties of the spin-resolved current completely characterize the QSH
state in this two-terminal geometry. As long as
$t_{\rm R} < t_{\rm SO}$ we remain in the QSH regime. 
For $t_{\rm R} \ge t_{\rm SO}$, the topological gap in bulk graphene 
vanishes and the system is no longer a QSH insulator.

\begin{figure}[t] \centering
\includegraphics[width=8cm,height=12cm,keepaspectratio]{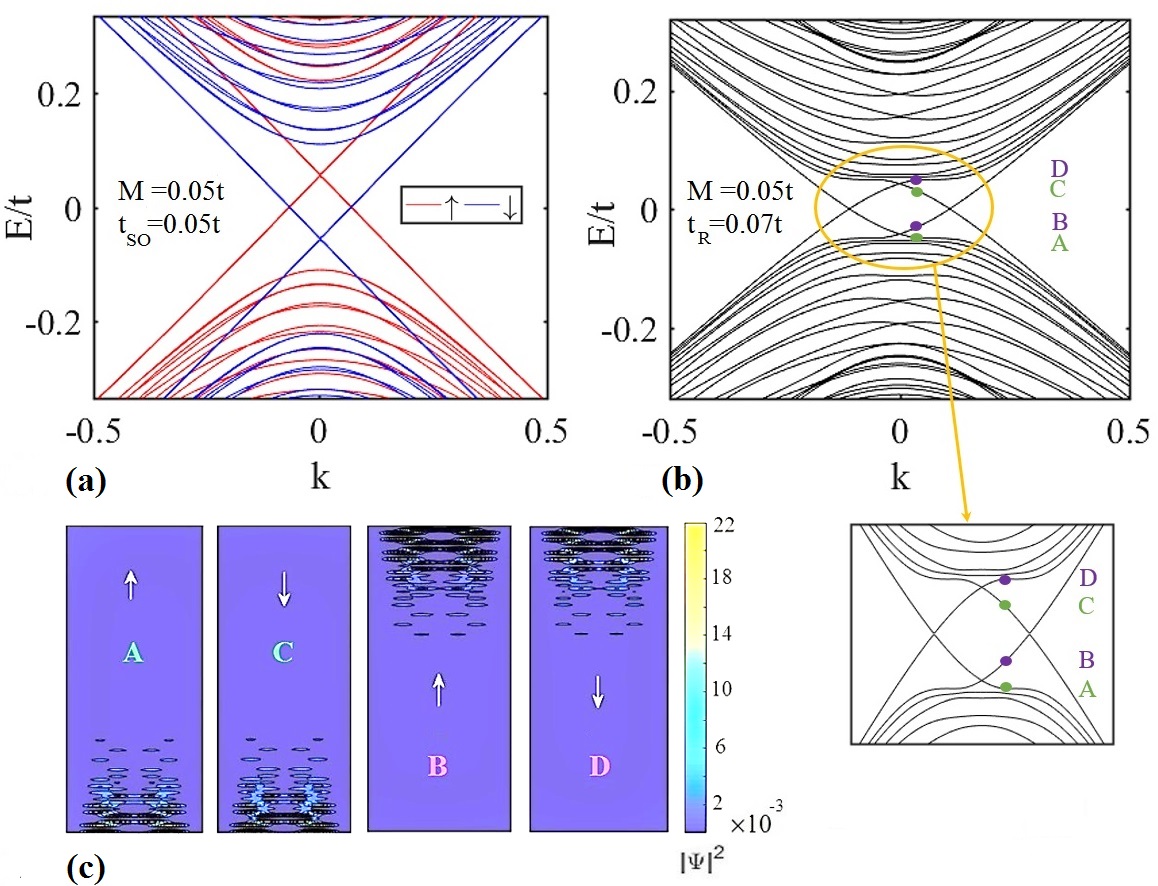}
\caption{\label{figurethree}
Band structures  and wave-function spatial distributions 
for a 196-AGNR in the presence of a FM
exchange field $M = 0.05t$.
(a) Band structure in the presence of intrinsic SOI  $t_{\rm SO} = 0.05t$.
(b) Band structure in the presence of Rashba SOI $t_{\rm R} = 0.07t$. A, B, C, D label the four
chiral edge states (two at each edge) with wave vector $k = 0.05$, see also the zoomed-in inset. (c) The wave-function spatial distributions in the nanoribbon unit cell shown in Fig.~\ref{figuretwo}(a)
for the four labeled edge states.} \end{figure}

\subsection{\label{sec:level32} Magnetic exchange fields: the QAH effect}

We now investigate the effect of adding a FM field that breaks
TRS and can trigger a
topological phase transition from the QSH to the QAH phase. In the absence of SOI (intrinsic and Rashba) a FM splits the two spin-degenerate
Dirac cones and Dirac points with respect to each other. 
As a result of this splitting, the valence band of
the spin-up states crosses the conduction band of the spin-down state at wave vectors $k = \pm k_0$.
Suppose that only an intrinsic SOI is now switched on.
Since this SOI does not couple opposite-spin states, it leaves these crossing points unaffected,
but it opens a gap at the two Dirac points. In the resulting energy gap around $E= 0$, two pairs of 
linearly-dispersed helical edge states now appear.  
These are precisely the four helical states
of Fig.~\ref{figureone}(b)  except that now the spin degeneracy 
is lifted by the FM exchange field, see Fig.~\ref{figurethree}(a).
As for the case of zero exchange field,
these four helical states give rise to the QSH effect, which therefore  occurs even in the absence of 
TRS, when the only SOI present is of the spin-conserving intrinsic type
\footnote{According to Ref.~\onlinecite{Yang2011}, 
there should be 
a transition to a QAH state for a critical value of a FM exchange field, 
even when $t_{\rm R} = 0$. 
We don't see this transition, 
since, when only intrinsic SOI is present, 
the FM field does not mix the spins and does not
change the eigenstates.}.

The situation is different when a Rashba SOI is switched on. 
Let us consider first the case where this
is the only SOI present ($t_{\rm SO}= 0$).
Now the SOI couples opposite spin states,
and therefore the level crossings between valence and conduction states,
brought about by the FM exchange field, are
replaced by avoided level crossings. The resulting bulk energy gap opening around $E= 0$ has a 
topological nature very different from the one of the QSH phase, and so is also the case for the four 
linearly-dispersed edge states appearing in this gap, shown in Fig.~\ref{figurethree}(b)  for 
a 196-AGNR. 
An analysis of these four branches
reveals that they are 1D {\it chiral} edge states: the two branches 
localized at the same edge move in the same direction. 
This is shown in Fig.~\ref{figurethree}(c) where we plot the wavefunction space distribution $|\Psi|^2$ 
inside the unit cell for four
states with the same $k = 0.05$, belonging to the four edge channels.
States A and C (B and D) belong to the two left(right)-mover branches and are localized at the bottom(tiop) edge.
The two chiral states localized at
the same edge have predominantly opposite spins, 
though the quantization is not perfect since spin is not conserved when Rashba SOI is present. 

\begin{figure}[t]\centering 
\includegraphics[width=8.7cm,height=10cm,keepaspectratio]{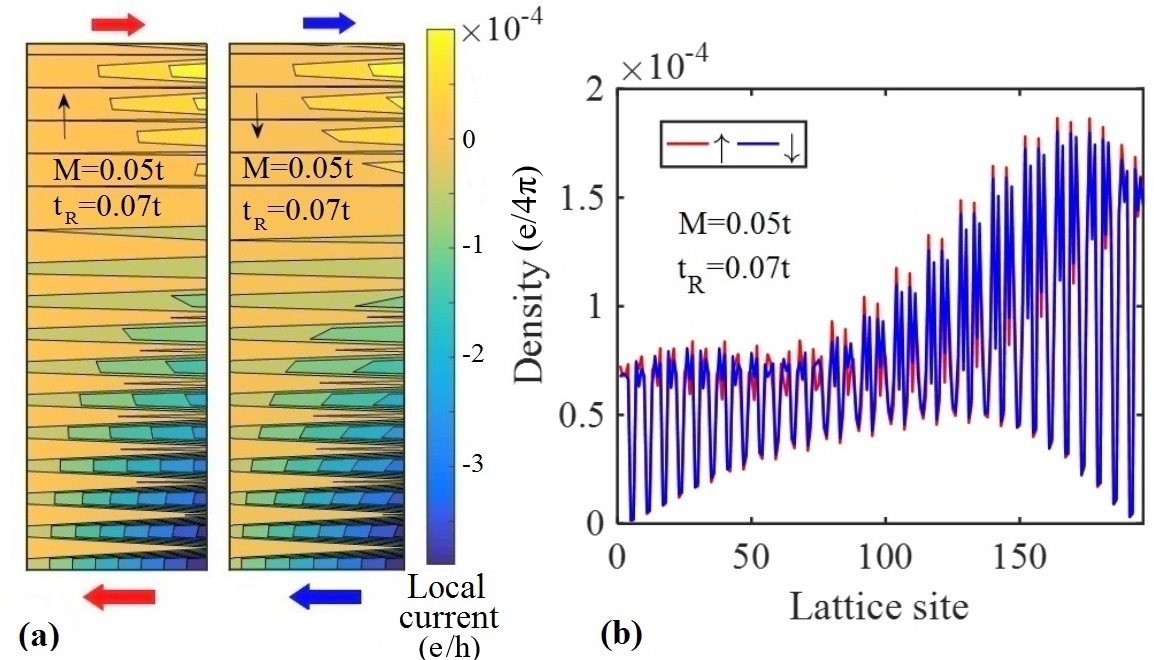}
\caption{\label{figurefour} (a) The local spin-resolved bond current and (b) the
non-equilibrium spin-resolved charge density for the 196-AGNR of Fig.~\ref{figurethree}
in the presence of Rashba SOI and FM exchange field.} \end{figure}

Based on the original edge-state approach to the quantum Hall effect introduced by B{\" u}ttiker \cite{Buttiker1988}, 
the presence of two pairs of conducting chiral states at the edges signals the onset of a QAH phase,
characterized in this case by a Chern number ${\cal C} =  2$, leading to a quantized Hall conductance $2 G_Q$~\cite{Yang2011}. This occurrence is directly visible in the non-equilibrium two-terminal transport calculations. 
In Fig.~\ref{figurefour}(a), we plot the spin-resolved bond (charge) current in the nanoribbon unit cell, calculated
using Eq.~\ref{eq:5}, for 
$eV<E_g$
[Fig.~\ref{figurethree}(b)].
We can clearly see that the net current, going in this case from right to left, is due to the 
two spin-polarized conducting chiral edge states
localized at the bottom edge, each contributing a quantized value to the Hall conductance.
These bottom (left-mover) edge states, being injected from the "source" electrode, will be at a higher chemical
potential than the top edge states, and therefore will carry more current. 
The additional current is proportional to the increase of the 
1D electron density at the top edge, as shown in Fig.~\ref{figurefour}(b), 
where we plot the non-equilibrium spin-resolved
density inside the nanoribbon unit cell.

\begin{figure}[b]\centering
\includegraphics[width=8cm,height=9cm,keepaspectratio]{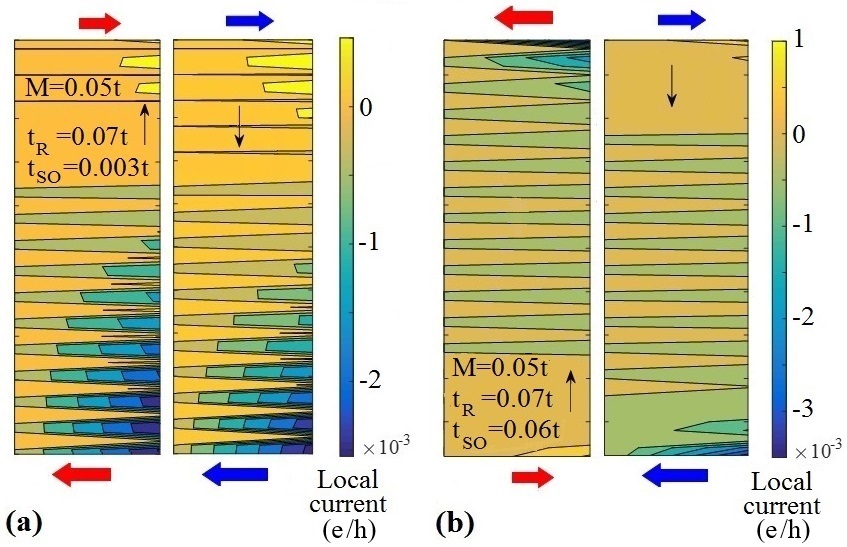}
\caption{\label{figurefive} Local spin-resolved bond current for the
same system as in Fig.~\ref{figurefour}, when an additional intrinsic SOI has been switched on.
(a) $t_{\rm SO} = 0.003 t \ll t_{\rm R} = 0.07 t$:  the system remains in the QAH regime; (b)  $t_{\rm SO} = 0.06 t \simeq   t_{\rm R} = 0.07 t$:
helical edge states indicate that the system is undergoing a phase transition to the QSH regime.} 
\end{figure}

When in addition 
an intrinsic SOI is 
added, 
the two topological phases, QAH and QSH, start to 
compete. According to the Chern number analysis of Ref.~\onlinecite{Yang2011}, when
the ratio $\lambda = t_{\rm SO}/t_{\rm R}$ becomes larger than a magnetization-dependent critical value, the system
undergoes a topological phase transition from a QAH phase to a QSH phase. This theoretical prediction is
essentially consistent with our transport calculations, with an important caveat. In Fig.~\ref{figurefive}   
we plot the local spin-resolved bond current for two values of $\lambda$ for $M = 0.05 t$.
As shown in Fig.~\ref{figurefive}(a) when $\lambda \ll 1$, the system
remains in the QAH phase. For $\lambda \gg 1$ (not shown), we find
that a transition to a QSH phase has taken place. 
However, in the intermediate regime $\lambda \simeq  1$ 
[Fig.~\ref{figurefive}(b)]
although transport via helical edge states characteristic of
the QSH phase is dominant, a non-zero bond current is present also
in the middle of the nanoribbon, possibly due to states
that are neither chiral non helical. This result might indicate that, for a finite-width nanoribbon, the transition
between the QAH and QSH phases as function of $\lambda $ occurs via a crossover regime rather than at a critical
value $\lambda_{\rm cr}$.

We conclude
by briefly discussing the effect of an AFM exchange field, which
when added to the Hamiltonian of Eq.~\ref{eq:1}, breaks both
TRS $\Theta$ and inversion symmetry $\Pi$, but
preserves the combination of the two. In particular, the AFM field preserves the symmetry 
$S \equiv \Theta T_{\Delta}$, where $T_{\Delta}$ is the lattice translational symmetry swapping the two
graphene sublattices. This occurrence is similar but not identical to the one considered in 
Ref.~\onlinecite{Mong2010}, where the combined symmetry involved the product of 
$\Theta$ and $T_{1/2}$, the lattice translational  
symmetry of the "primitive" structural lattice broken by the AFM order. For this case the system is a topological phase called AFM TI, 
which shares some properties of the "strong" topological insulators.
On the other hand, our 
calculations for an AGNR in the presence of AFM exchange and Rashba SOI show that  for $t_{\rm R}/M < 1$ the spectrum has a bulk energy gap centered at $E= 0$, and the system is a trivial insulator. At $t_{\rm R}/M \simeq M$ the gap closes with flat bands at $E= 0$ and opens again for $t_{\rm R}/ M > 1$. 
However, no topological chiral edge states ever emerge in such a gap and the system remains a trivial insulator. 
Thus, 
in the presence of an AFM 
TRS-breaking field,
the system can only display a QSH phase, which happens
when $t_{\rm SO} > t_{\rm R}$. 

\section{\label{sec:level4}CONCLUSIONS} 
We have carried out non-equilibrium quantum transport calculations on armchair graphene nanoribbons
in a two-terminal geometry in the presence of spin-orbit interactions (SOIs) and 
TRS-breaking
exchange fields. Depending on the relative strength of these interactions,
the system can sustain both a quantum spin Hall (QSH) and a quantum anomalous Hall (QAH) regime.
Both phases are completely characterized  by the nature of their conductive topological edge states, 
which control the electronic transport and display specific features in the spin-resolved local current
and out-of-equilibrium charge density.  
In the presence of a FM exchange field, 
it is possible to trigger a topological phase transition
from a QAH phase to a QSH phase by changing the relative
strength of the intrinsic and Rashba SOI.
Therefore the QSH phase can survive even in the absence of 
TRS.

\section{Acknowledgments}
This work was supported by the Faculty of Technology and by the Department of
Physics and Electrical Engineering at Linnaeus University (Sweden).  We
acknowledge financial support from the Swedish Research Council (VR) through
Grant No.~621-2014-4785, and by the Carl Tryggers Stiftelse through Grant No.~CTS 14:178.  Computational resources have been provided by the Lunarc Center for
Scientific and Technical Computing at Lund University.  CMC acknowledges
valuable interactions with A.H. MacDonald.

\bibliography{REF} \end{document}